\documentclass[aps, pra, showpacs,showkeys,twocolumn, superscriptaddress, longbibliography]{revtex4-1} %,twocolumn
\usepackage{graphicx,amsmath,hyperref}
\usepackage{xcolor}
\usepackage{braket}
\usepackage{amsmath, amssymb}
\usepackage{caption}
\usepackage{subcaption}
\usepackage{natbib}

\newcommand{\fla}[1]{\begin{flalign}#1\end{flalign}}

\newcommand{\beq}{\begin{equation}}
\newcommand{\eeq}{\end{equation}}
\newcommand{\ba}{\begin{eqnarray}}
\newcommand{\ea}{\end{eqnarray}}
\newcommand{\bea}{\begin{eqnarray}}
\newcommand{\eea}{\end{eqnarray}}
\newcommand{\bma}{\begin{subequations}}
\newcommand{\ema}{\end{subequations}}
\newcommand{\bwt}{\begin{widetext}}
\newcommand{\ewt}{\end{widetext}}

\definecolor{green1}{RGB}{0, 110, 61}
\definecolor{brown1}{RGB}{212, 55, 0}
\definecolor{blue1}{RGB}{34, 65, 156}
\definecolor{red1}{RGB}{239, 67, 63}
\newcommand{\RE}{\mathrm{Re}}
\newcommand{\IM}{\mathrm{Im}}
\def\abs#1{|\,#1\,|}

\begin{document}

\title{Toward highly efficient multimode superconducting quantum memory}

\author{Aleksei R. Matanin}
\affiliation{FMN Laboratory, Bauman Moscow State Technical University, Moscow 105005, Russia}
\affiliation{Dukhov Automatics Research Institute (VNIIA), Moscow 127030, Russia}

\author{Konstantin I. Gerasimov}
\author{Eugene S. Moiseev}
\affiliation{Kazan Quantum Center, Kazan National Research Technical University, Kazan,  420111, Russia}

\author{Nikita S. Smirnov}
\author{Anton I. Ivanov}
\author{Elizaveta I. Malevannaya}
\affiliation{FMN Laboratory, Bauman Moscow State Technical University, Moscow 105005, Russia}
\affiliation{Dukhov Automatics Research Institute (VNIIA), Moscow 127030, Russia}
\author{Victor I. Polozov}
\author{Eugeny V. Zikiy}
\author{Andrey A. Samoilov}
\affiliation{FMN Laboratory, Bauman Moscow State Technical University, Moscow 105005, Russia}
%\affiliation{Dukhov Automatics Research Institute (VNIIA), Moscow 127030, Russia}

%\author{Eugeny V. Zikiy}
%\affiliation{Dukhov Automatics Research Institute (VNIIA), Moscow 127030, Russia}

%\author{Andrey A. Samoilov}

\author{Ilya A. Rodionov}
\thanks{irodionov@bmstu.ru}
\affiliation{FMN Laboratory, Bauman Moscow State Technical University, Moscow 105005, Russia}
\affiliation{Dukhov Automatics Research Institute (VNIIA), Moscow 127030, Russia}

\author{Sergey A. Moiseev}
\thanks{s.a.moiseev@kazanqc.org}
\affiliation{Kazan Quantum Center, Kazan National Research Technical University, Kazan, 420111, Russia}

\begin{abstract}
We experimentally demonstrate a microwave quantum storage for two spectral modes of microwave radiation in on-chip system of eight coplanar superconducting resonators. Single mode storage shows a power efficiency of up to $60\pm3\%$ at single photon energy and more than $73\pm3\%$ at higher intensity. The noiseless character of the storage is confirmed by coherent state quantum process tomography. The demonstrated efficiency is an order of magnitude higher than the previously reported data for multimode microwave quantum memory. The proposed on-chip quantum memory architecture can be easily integrated into the state-of-the-art superconducting quantum circuits technology without any coherence compromises. The obtained results are in good agreement with the proposed theory, which pave the way for building a practical multimode microwave memory for superconducting quantum circuits.
\end{abstract}

%Microwave quantum memory promises advanced  capabilities for noisy intermediate-scale superconducting quantum computers. Existing approaches to microwave quantum memory lack complete combination of high efficiency, long storage time, noiselessness and multi-qubit capacity. Here we report an efficient microwave broadband multimode quantum memory. The memory stores two spectral modes of single photon level microwave radiation in on-chip system of eight coplanar superconducting resonators.Single mode storage shows a power efficiency of up to $60\pm 3\%$ at single photon energy and more than $73\pm 3\%$ at higher intensity. The demonstrated efficiency is an order of magnitude larger than the previously reported multimode microwave quantum memory. The noiseless character of the storage is confirmed by coherent state quantum process tomography. The demonstrated results pave the way to further increase in efficiency and hence building a practical multimode microwave memory for superconducting quantum circuits.

\maketitle
\section{Introduction}
Fault-tolerant quantum computing and quantum internet require quantum memory as an essential building block of a future quantum information processing platform \cite{kimble2008, Lvovsky2009, Wehner2018, Blais2021, Matteo2011}. Superconducting circuits quantum electrodynamics (cQED) is among the leading realizations of intermediate-scale quantum computers \cite{Morten2020, Devoret2013}.
Meanwhile there is a strong motivation to break the wall of nearest-neighbor qubit coupling using enhanced cQED architecture with integrated quantum memory \cite{Matteo2011, 2017-Naik-NatureComm, Pfaff2017, Axline2018,  Morten2020}. 
Moreover, it would allow to extend limited coherence time of the superconducting qubits, implement complex quantum algorithms \cite{QSearch-Lloyd-PRL-2008} and hardware-efficient quantum error correction \cite{Leghtas2013, Corcoles2015, Ofek2016, Rosenblum2018}. 
Compared with traditional superconducting qubits, high quality factor resonators have a superior potential for quantum state storage due to their impressive lifetime \cite{Ofek2016, Reagor2016, Wenner2014, Kobe2017}, efficient thermalization, no extra fridge control lines and ability to couple multiple qubits
\cite{2017-Naik-NatureComm, Kubo2011}.

Quantum memory based on a single superconducting resonator demonstrates high efficiency in storing a microwave photon with an optimal temporal mode \cite{Wenner2014,Flurin2015}.
However, the specific exponential rising mode of that memory complicates interconnections with other circuits. 
The memory stores only a single qubit with a fixed bandwidth that is limited by a coupling with an input waveguide.
To increase bandwidth of a memory and its multi-qubit capacity, it is promising to use multi-resonator schemes \cite{McKay2015, 2017-Naik-NatureComm, Moiseev_2016,Moiseev2017,2018-Moiseev-SR}, which can overcome the restrictions of a single resonator.

The first approach uses a linear chain of coupled resonators \cite{2017-Naik-NatureComm}. Interaction between resonators leads to an emergence of collective modes with different frequencies and different coupling constants between the resonators and closely spaced qubits.
Each of the collective modes is used to selectively store the quantum state of a particular qubit by tuning the qubit's frequency into resonance with that mode. 
However, this scheme provides efficient storage only for specific temporal waveform as in single-resonator quantum memory.

The multi-resonator approach \cite{2018-Moiseev-SR} exploits the ideas of photon/spin echo \cite{Moiseev2001, Riedmatten2008,
Tittel2010,
Grezes2015,
Ranjan2020,
Moiseev_2021} 
in a system of resonators with a linear periodic spacing of their resonant frequencies. An echo forms in such resonators and reemits an input pulse after $\tau=1/\Delta$, where $\Delta $ is the frequency spacing between resonators.
In this case, the effective bandwidth of the memory is  determined by the span of the formed frequency comb, which significantly exceeds the linewidth of an individual resonator. 
This idea was used for on demand storage of two temporal microwave modes at single-photon level in a system of 4 planar superconducting resonators \cite{2021-Bao-PRL}. The efficiency of 6\% was demonstrated in the work that turned out to be lower than the efficiency of 16.3\% achieved at room temperature on a system of 3D resonators with lower Q-factor \cite{2018-Moiseev-SR}. 
It raised a concern of implementing a highly efficient broadband multi-resonator memory on a system of planar resonators, since a direct interaction between the resonators may ruin the impedance matching.
Multimode microwave storage is also elaborated in spin ensembles with long-lived coherence   by echo-based protocols \cite{Julsgaard13,Afzelius_2013}.
However, the demonstrated results 
suffer from low efficiency $(<1\%)$ and considerable noise \cite{Grezes2015,Ranjan2020}.

\begin{figure*}
    \centering
    \includegraphics[width=\linewidth]{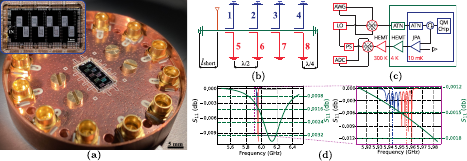}% fig1_main_v1PS Fig-1N.pdf
        \caption{\textbf{(a)} Image of the fabricated quantum memory device.
        The color area around the internal $\lambda/4$ resonators is a vortex-pinning hole array that  helps achieve high internal Q-factor by magnetic vortices trapping \cite{McRae2020}.
        Inset: optical micrograph of the quantum memory chip. “IN” indicates the input port that is connected to the
common resonator with a voltage tap.
\textbf{(b)} Principal scheme of the quantum memory device with the relative positions of the common resonator (\textcolor{green1}{green} line), first group internal resonators (\textcolor{blue1}{blue} line), second group of internal resonators (\textcolor{red1}{red} line) and input port (\textcolor{brown1}{brown} line).
\textbf{(c)} Simplified scheme of the experimental setup: arbitrary waveform generator (AWG), analog-to-digital converter (ADC), continuous-wave local oscillator (LO), chain of attenuators (ATN), quantum-limited Josephson parametric amplifier  (JPA), two low-noise amplifiers based on high-electron-mobility transistor (HEMT), phase shifter (PS). The colors indicate the different temperature stages for the components. \textbf{(d)} Simulated stationary reflection spectrum $S_{11}$ for the first (\textcolor{blue1}{blue} solid line) and the second (\textcolor{red1}{red} solid line) quantum memories and the frequency-shifted single common resonator (\textcolor{green1}{green} solid line). The carrier frequencies for first and second memory cells are indicated as dashed \textcolor{blue1}{blue} and \textcolor{red1}{red} vertical lines, respectively.} 
\label{fig:fig1}
\end{figure*}

Here, we demonstrate a broadband two mode microwave quantum storage on a system of planar resonators with storage efficiency of $60\pm3\%$ at single photon energy with fixed retrieval time $\tau=1/\Delta$.
%Here, we demonstrate a broadband two mode on-chip microwave quantum memory with storage efficiency of $60\pm3\%$ at single photon energy.
First, we portray our design, its underlying physical principles and show storage results of coherent pulses for two modes with independent carrier frequencies. 
Next, we present a complete quantum characterization of the storage for a single mode by quantum process tomography. 
Finally, the obtained results and future perspectives of unity efficiency, on-demand and long-lived storage for the realized quantum storage device are discussed.

\section{Results}

Our quantum storage device (Fig. \ref{fig:fig1}a) consists of one common resonator that interacts with eight internal superconducting resonators and coupling waveguide as it is depicted at principal scheme in Fig. \ref{fig:fig1}b. 
The common resonator has a designed frequency of 6 GHz with length of $8\cdot\lambda/4$, where $\lambda$ is the resonant wavelength. 
The waveguide is coupled to the common resonator with a coupling constant $\kappa\approx 281$ MHz.
Each of the eight internal resonators has a length of $\sim\lambda/4$ and is coupled to the common resonator with coupling constant $g\approx 12$ MHz. 
The internal resonators have unloaded Q-factor $\sim$ 5$\cdot$10$^{5}$ that corresponds to decay constant  $\gamma_n \approx $ 6 kHz that is same as decay constant for common resonator  
$\gamma_0 \approx $ 6 kHz.
The proper choice of $\kappa$, $g$, $\gamma_n$ and resonator frequencies 
%of $g$, $\kappa$ and $\gamma_n$
is crucial for impedance matching condition (see Appendix \ref{app:A}), that assures the efficient transfer of an input pulse from the waveguide into the internal resonators. 
We found that closely spaced planar resonators may directly interact with each other not only through the common resonator mode, exerting an unpredictable shift on their frequencies and breaking an impedance matching condition. 
We solve this problem by choosing the appropriate frequencies and relative position of the nearest resonators as shown in Fig. \ref{fig:fig1} \textbf{a}-\textbf{b}. 
Moreover, strong coupling ($g>\Delta$) and large Q-values of internal resonators lead to weak contrast in the classical measurement of continuous wave reflection, which are difficult to measure experimentally.
For example, the reflection contrast at these calculated parameters is on the order of -0.01 dB, as theoretically shown in Fig. \ref{fig:fig1} \textbf{d}.
Hence, we extract the experimental values of  $g, \kappa, \gamma$ by fitting the time domain data to our theoretical model (see Appendix \ref{app:A}).

The designed frequencies of the internal resonators range from 5.9895 GHz to 6.0105 GHz with a step of $\Delta=3$ MHz.
We keep the position of each internal resonator in the voltage antinodes of the common resonator as it is shown in Fig. \ref{fig:fig1}b. 
Such positioning allows to choose an identical coupling constant for all internal resonators since coupling depends only on a mutual capacitance that is fabricated with good precision.
In turn, the position $l_{\text{short}}$ of the input port relatively to the common resonator is crucial for the value of coupling constant $\kappa$ (see Appendix \ref{app:B}).

We adopt a small difference in $\Delta$ of $0.5$ MHz  for two groups of resonators labeled as 1-4 and 5-8 in Fig. \ref{fig:fig1}b to select two frequency modes for storage and separate them in time domain.  
We measure the resonators 1 to 4 to be frequency spaced by $\Delta_{1-4}=3.55$ MHz, while resonators 5 to 8 are spaced by $\Delta_{5-8}=3.08$ MHz.
Hence we consider these two groups of resonators as two independent quantum memory cells, namely first and second, for two modes with carrier frequencies 5.9436 GHZ and 5.9549 GHz, respectively.  It must be noted that our device stores a two-mode quantum state for a fixed time, and hence it is not full on-demand quantum memory. However, for the sake of simplicity, we will call these two groups of resonators quantum memory cells.

Simplified scheme of the experimental setup for characterizing quantum storage device is depicted in Fig. \ref{fig:fig1}c (see also Appendix \ref{app:C} for more details).
The quantum device is installed in a dilution refrigerator at the base stage with temperature of 10 mK together with microwave circulators and a quantum-limited Josephson parametric amplifier. An intermediate frequency (up to 500 MHz) signal from an arbitrary waveform generator is mixed with a radio frequency signal from the continuous wave local oscillator on an IQ mixer. 
The mixing result is used to prepare pulses with fixed amplitude, phase and carrier frequency shifted relative to the local oscillator frequency by an intermediate frequency.
The pulses are then attenuated by the total -80 dB that prepares the pulses in a coherent state with the noise temperature being close to the base stage temperature of 10 mK \cite{Paris96}. According to this attenuation value the total signal power in single photon regime is $\sim$-150 dBm.
The circulator directs the prepared coherent state of microwave radiation into the quantum memory chip.
The reflected signal from the memory cell is routed by the circulator into a quantum-limited Josephson parametric amplifier and two low-noise amplifiers.
These two amplifiers are based on high-electron-mobility transistor (HEMT) and placed at temperature of 4K and room temperature, respectively.
%Hence the memory's output is totally amplified by $\sim$90 dB that brings noise of it's quantum fluctuation 20 dB above noise level of the analog-to-digital converter.
The amplified signal is mixed on the IQ mixer with the local oscillator that is phased locked to an input signal and phase-tuned to measure arbitrary quadrature.
The resulted quadratures in I and Q channels are sampled by fast analog-to-digital converters for further analysis.

First we present experimental results on storage efficiency for intensive coherent microwave pulses. 
For each group of resonators, we perform independent experiments by tuning the carrier frequency of the input pulse to the central frequency of each group of the resonators, as shown in Fig. \ref{fig:fig1}d.
The input pulse has a Gaussian waveform with full width at half maximum of 115 ns that matches the bandwidth of each memory cell. 
The pulses are measured by heterodyne detection with local oscillator being detuned by 50 MHz.

An amplitude of the resulted beating signal in time domain is compared to the one of reference pulse with the same amplitude, but being sent out of resonance with the quantum storage device.
The reference pulse is detuned from the resonances of both quantum memories.
However, the intensity of the reference pulse depends on the choice of the detuning as HEMT amplifier's gain varies over frequency.
We determine the average value of pulse intensity at 21 frequencies in the range from 5.8 GHz to 5.9 GHz. 
The measurements show that the variance of the reflected signal intensity is 10.7$\%$. 
This value is used to determine an accuracy in calculating the efficiency of the quantum storage device, whereas the decrease in signal to noise ratio and the corresponding error   we compensate by a proportional larger averaging (up to $5\cdot10^7$ in single photon regime).

\begin{figure*}
\centering
\includegraphics[width=\linewidth]{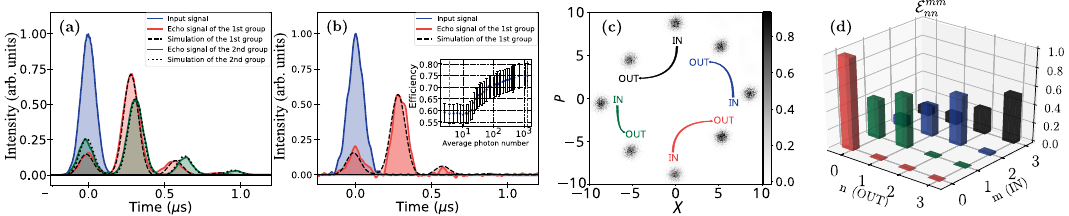}
\caption{
\textbf{(a)} The signals from the first (\textcolor{red1}{red} solid line) and second (\textcolor{green1}{green} solid line) memory cells at high power with large average photon number.
\textbf{(b)} The signals from the first memory cell (\textcolor{red1}{red} solid line) at single photon level input pulse intensities.
The signals are normalized to the maximum intensity of the input Gaussian-shaped coherent pulse (\textcolor{blue1}{blue} solid line).
The black dashed and dotted lines are theoretical simulations with parameters discussed in the text. Inset: Efficiency dependence for average photon number. \textcolor{blue1}{Blue} dashed line corresponds to data in fig. 2a and \textcolor{red1}{red} dashed line corresponds to data in fig. 2b.
\textbf{(c)} Four pairs of experimentally acquired phase space distribution for input pulses ('IN') and the corresponding memory's response ('OUT'). Single color and arrow indicate a single pair of phase space distributions of conjugate position $X$ and momentum $P$ for the state of the mode being sent into the memory and recalled state. \textbf{(d)} Diagonal elements of the reconstructed quantum process tensor.
}
\label{fig:fig2}
\end{figure*}

In Fig. \ref{fig:fig2}a we present normalized time-domain data at the outputs of both memories for a input pulse with large average photon number ($n_{ph}\gg 1$).
The first memory cell with  resonators 1-4 irradiates $75\pm8\%$ of the input power in the first echo after 277 ns, while the second memory cell (5-8) outputs $52\pm5.7\%$ of input power in its first echo after 310 ns. 
At the same time, the reflected power with no delay is larger for the second memory cell. 
We attribute this difference in efficiencies to the impedance matching condition being better fulfilled in the first memory cell than in the second one due to common resonator frequency shift. 
Fig. \ref{fig:fig1}d shows theoretically simulated reflection spectrum of the both memory cells with respect to the common resonator. Here in the regime of the detuned common resonator, the effective coupling monotonically decreases for the internal resonators closer to the frequency of the common resonator.  Overall, it leads to a smaller efficiency of the second memory cell in the absence of multi-resonator impedance matching. 
For both memory cells, the total integrated power over 1.5 $\mu$s was $\sim$96$\%$ of the input energy.
The acquired experimental data for both memories is fitted to our theoretical model with a good agreement. 
In principle, the different frequency spacing of two group of resonators can be chosen. 
For example, for ratio of $\Delta_{1-4}/\Delta_{5-8}=2$ both frequency modes can be efficiently  stored with insignificant overlap in time domain (see Appendix \ref{app:A}).

Next we perform experiments on a storage of microwave pulses at a single photon level in the first quantum memory cell. 
The number of photon is calibrated by the amplitude moments method \cite{Eichler2011}.
The measured amplitude moments together with the known gain and noise temperature of the HEMT-based amplifier allow us to determine the probe’s amplitude at the input of the amplifier using the reconstructing technique from \cite{Eichler2011}.  Taking into account the negligible loss in the circulator ($<$0.2 dB), we assume the reconstructed amplitude is an amplitude of the probe being sent to the memory unit.

We gradually decrease the intensity of the signal pulse and observe a reduction in the efficiency from $75\pm8\%$ to $58\pm6.4\%$ in single photon regime.
The inset of Fig. \ref{fig:fig2}b shows a dependence of the efficiency on the average photon number. 
It is an expected effect that is associated with  saturable two-level system (TLS) defects in the superconducting resonators \cite{2008-Gao-APL,2017-Brehm-APL,2020-Kudra-APL}.
We model this effect by increasing the decay rate to $\gamma_n =165 \;\text{kHz}$ in single photon regime, thus effectively reducing Q-factor to 36$\cdot$10$^{3}$. 
Fig. \ref{fig:fig2}b shows time domain data on amplitude of the echo averaged over a series of experiments with single photon level coherent pulses. This data is fitted to the theoretical model, that includes the change in the decay constants of the resonator modes.

We confirm a noiseless character of our quantum storage by performing coherent state quantum process tomography on our memory  with a single microwave mode \cite{Lobino08}. We specify the quantum storage operation as a process tensor $\mathcal{E}^{nm}_{jk}$ \cite{Lobino09} that linearly relates input $\rho^{\text{in}}$ to output $\rho^{\text{out}}$ density matrices of the microwave mode in Fock basis:
\fla{
\rho^{\text{out}}_{jk} = \sum^{H}_{m,n=0} \mathcal{E}^{nm}_{jk} \rho^{\text{in}}_{nm}, 
}
where $H$ is high-energy cut-off that truncates the Fock space of interest.   
If the process is phase invariant with respect to phase of an input coherent state, then only diagonal elements of the process tensor are non-trivial \cite{Rahimi_Keshari_2011}.

First, we demonstrate that our quantum memory is phase invariant by consequently  storing coherent states $\rho^{\text{in}} = \ket{\alpha e^{i\phi} }\bra{\alpha e^{i\phi}}$ with the same amplitude ($\abs{\alpha}\sim 9$) but with different phases. 
The input and echo pulses in Gaussian temporal mode are measured by homodyne detection (see Appendix \ref{app:D}). 
Based on these measurements, we reconstruct phase-space distributions of quadratures for input and echo pulses.
The memory shows a constant phase shift of $\sim \pi/4$ radians for coherent states with different phases as it is depicted in Fig. \ref{fig:fig2}c. 
The same phase-invariant behaviour is observed at lower amplitudes ($\abs{\alpha} \sim 1$).   

Next, we reconstruct the process tensor by the maximum-likelihood method \cite{Anis_2012}. 
We fix a phase of the input pulse and vary its amplitude from 0 to 1.2 with a step of 0.02. 
For each amplitude's value we perform multiple experiments and  acquire $2\cdot10^5$ quadratures. The sampled distribution of recorded quadratures is used to reconstruct the process tensor in the Fock space of dimension 4, as it is depicted in Fig. \ref{fig:fig2}d.  The presented diagonal elements $\mathcal{E}^{mm}_{nn}$ correspond to probabilities of detecting n-th Fock state at the output of the quantum memory for m-th Fock state being sent as an input. 

The storage proves to be noiseless with a probability of getting a noisy photon less than 1\%. From the reconstructed process tensor, the single photon power efficiency is estimated to be $60\pm3\%$  that is consistent with heterodyne measurements at low intensity. The non-linear trend of power efficiency is presented with $73 \pm 3\%$ being a reconstructed efficiency for $\ket{2}$ and $\ket{3}$ Fock states. 
As the single pulse is stored in the group of 4 resonator we expect that the observed saturation of the TLS at low intensity may be attributed to the relatively long life-time of the TLS compare to the period of the tomographic experiment ($\sim$50 $\mu$s) that may be a subject of separate study. %This effect will be a subject of a separate study. 

\section{Discussion}
The demonstrated efficiencies of 60\% for single photons and 75\% for higher intensities in the single quantum memory cell can be increased if the condition of impedance matching is exactly met.
Our theoretical model predicts efficiency of $73\%$ for single photons and more than 96\% for multi-photon states under complete impedance matching with the efficiency being limited only by the coherence time of the internal resonators. 
Moreover, further technological improvements in term of TLS mitigation with different superconductive materials like niobium and tantalum \cite{Place2021,Wang2022} can significantly reduce the losses in the internal resonators and increase efficiency even closer to unity in single photon regime with a lifetime about a millisecond.

It is worth noting that the demonstrated quantum storage suppresses coupling of the quantum noise from the internal resonator into output waveguide according to our model.
We show (see Appendix \ref{app:A}) that the suppression factors for $\gamma_{0}$ = $\gamma_{n}$ = 165 kHz and 6 kHz are $\sim$0.14 and 0.006, respectively. 
It means that the quantum memory can operate at temperatures higher than 10 mK. 
For example, single photon storage with a signal-to-noise ratio of $\sim$100 can be achieved at temperatures of 100 mK and 600 mK for $\gamma_{0}$ = $\gamma_{n}$ = 165 kHz and 6 kHz, respectively. 
However, we doubt that the effect can be observed with aluminium resonators due to aluminium's relatively low critical temperature and the decrease of the internal resonator's quality factor at the mentioned temperatures. The effect may be observed with materials having a higher critical temperature, such as niobium, titanium nitride, tantalum, etc. 
%The proposed designed may be also valuable for optical quantum storage at room temperature in a finite array of whispering gallery mode resonators \cite{Moiseev_2016}. 

The demonstrated chip stores a signal pulse for a fixed time interval, so it plays the role of a delay line for now. 
However, the scheme can be modified for on-demand storage, if a fast switch \cite{Flurin2015} is used to connect and disconnect the quantum memory from the input waveguide.
In this way, the storage time of signal pulses is a multiple of $n /\Delta$, where $n$ is an arbitrary integer that is set by the switch.
Recently, it was shown that one switch is enough to achieve on-demand storage with high efficiency and low noise at optimal parameters \cite{Moiseev2020}. 
In comparison with single resonator memories \cite{Wenner2014,Flurin2015}, the switch may operate when the internal resonators are excited with the common resonator being empty that avoids corruption of the stored data from the switch's noise. 
The switch induced losses in the common resonator would not significantly effect the performance of the quantum memory (see Appendix \ref{app:A}).
The memory design with identical $\Delta$ would be able to store multiple temporal qubits.
This multi-qubit memory in conjunction with a quantum router would allow to implement quantum random access memory \cite{QRAM-Lloyd-PRL-2008, MoiseevJMO2016,Chen2021}, where the role of the router can be realized by superconducting circuits with Josephson junctions.

In conclusion, we presented noiseless two-mode on-chip quantum storage for microwave photons that is compatible with cQED architecture. Among different photon storage experiments in frequency-comb media, such as rare-earth ions for optical storage \cite{Sabooni2013} and nuclear transitions for gamma-photon storage \cite{Zhang2019}, the presented results demonstrate the highest efficiency. 
The obtained experimental results agree well with the theory, which predicts a further considerable increase of efficiency towards 100\%.
It will open an avenue for a practical multi-qubit memory with arbitrary temporal and spectral mode multiplexing with applications in superconducting quantum processing.

%\section{Data availability}
%The data that support the findings of this study are available from the corresponding
%author upon reasonable request.

\begin{acknowledgments}
The device was fabricated at the BMSTU Nanofabrication Facility (Functional Micro/Nanosystems, FMNS REC, ID 74300).
The work of S.A.M., K.I.G and E.S.M. is carried out with financial support of the Ministry of Education and Science of Russia, Reg. number NIOKRT 121020400113-1. 
Authors thank Tatiana G. Konstantinova for assistance in manufacturing of the quantum memory chip and Vladimir V. Echeistov for the help in configuring the room temperature electronics. 
S.A.M. acknowledges Nikolay S. Perminov for useful discussions,  E.S.M. thanks Arina Y. Tashchilina for helpful discussions. 
\end{acknowledgments}

\appendix
\section{Theoretical model} \label{app:A}

The dynamics of the memory is governed by Heisenberg-Langevin equations for the system of the $N$ internal resonators and one common resonator that is coupled to an input waveguide \cite{2018-Moiseev-SR}: 
\fla{
\label{eq1}
&[\partial_{t}+i\Delta_n+\frac{\gamma_n}{2}]\hat{b}_n+ig_n\hat{a}_c=\sqrt{\gamma_n}\hat{F}_{n}, 
\\ 
\label{eq2}
&[\partial_{t}+i\Delta_0 +\frac{\kappa+\gamma_0}{2}]\hat{a}_c+i\sum_n g_n \hat{b}_n=\sqrt{\kappa}\hat{a}_{in}+\sqrt{\gamma_0}\hat{F}_{c},
}
where $\hat{b}_n$ and $\hat{a}_c$ are annihilation operators for mode of n-th internal and common resonators,  respectively, with commutation relationship $\left[ b_{n}, b^{\dagger}_{m} \right] = \delta_{n,m} $ and $\left[ b_{n}, b^{\dagger}_{m} \right] = \delta_{n,m} $ ; $g_n$ is coupling constant between the common resonator and n-th internal resonator;  $\Delta_n$ and $\Delta_0$ are the detunings of the n-th resonator and common resonator, respectively, from the carrier frequency of the input field that is used as a rotating frame. 
The coupling between the common resonator and the input waveguide with coupling constant $\kappa$ is described as input-output relations \cite{Gardiner1985}:
\fla{
\hat{a}_{in}(t)+\hat{a}_{out}(t)=\sqrt{\kappa}\hat{a}_c(t), \label{eq::in-out-time}
}
with $\hat{a}_{\text{in}}(t)$ ($\hat{a}_{\text{out}}(t)$) being the annihilation operator of the photon in the input (output) waveguide with commutation relationship $\left[ \hat{a}_{\text{in}(\text{out})}(t),\hat{a}_{\text{in}(\text{out})}^{\dagger}(t') \right] =\delta(t-t')$.
The incoherent decay of the n-th internal resonator with the rate $\gamma_n$ and common resonator with the rate $\gamma_0$ are accompanied by Langevin operators $\hat{F}_{n}(t)$ and $\hat{F}_{c}(t)$, respectively with commutation rules $[\hat{F}_{n}(t),\hat{F}_{m}^{\dagger}(t')]=\delta_{n,m}\delta(t-t')$ and 
$[\hat{F}_{c}(t),\hat{F}_{c}^{\dagger}(t')]=\delta(t-t')$.

We solve equations (\ref{eq1}-\ref{eq2}) by applying  Fourier transformation such that
\fla{
\hat{A}_n(t)=\frac{1}{\sqrt{2\pi}}\int d\omega \hat{A}_n(\omega)e^{-i\omega t},
}
where $\hat{A}_n(t)$ is the field or the noise operator of interest in time domain, $\hat{A}_n(\omega)$ is its Fourier image that satisfies commutation relations
 $[\hat{A}_n(\omega),\hat{A}_m^{\dagger}(\omega')]=\delta_{n,m}\delta (\omega-\omega')$.
After the transformation the obtained algebraic system of equation is solved to find the Fourier image of the common resonator:
\fla{
& \hat{a}_c (\omega)=2\frac{\sqrt{\kappa}\hat{a}_{\text{in}} (\omega)+\sqrt{\gamma_0}\hat{F}_{\Sigma}(\omega)}{\kappa+\gamma_0-2i(\omega-\Delta_0)+\chi (\omega)},
\label{eq::cr}
}
where $\chi(\omega)= \RE \chi (\omega)+i\cdot\IM\chi (\omega) $ is an effective permittivity of the memory with real and imaginary parts: 
\fla{
\RE \chi (\omega) =\sum_{n=1}^{N}\frac{g_n^{2}\gamma_{n}}{(\gamma_n/2)^{2}+(\Delta_{n}-\omega)^{2}}, \\
\IM\chi (\omega) =\sum_{n=1}^{N}\frac{g_n^{2}(\omega-\Delta_{n})}{(\gamma_n/2)^{2}+(\Delta_{n}-\omega)^{2}}.
}
The solution \eqref{eq::cr} contains an effective Langevin operator $\hat{F}_{\Sigma}(\omega)$
\fla{
\hat{F}_{\Sigma}(\omega)=\hat{F}_{c}(\omega)-i\sum^{N}_{n=1}\frac{g_n\sqrt{\gamma_n/\gamma_0}}{\gamma_n/2+i(\Delta_n-\omega)} \hat{F}_{n}(\omega)
}
that is essential for determining noise of the quantum memory at finite temperatures.

We obtain the Fourier image of the memory output by combing the solution \eqref{eq::cr} and input-output relationship for the common resonator \eqref{eq::in-out}:
%\fla{
%\hat{a}_{out}(\omega)=S(\omega)\cdot \hat{a}_{in}(\omega)+ M(\omega) \hat{F}_{\Sigma}(\omega),
%\label{eq::fs}}

\begin{eqnarray}\label{eq::in-out}
& \hat{a}_{out}(\omega)=S(\omega)\cdot \hat{a}_{in}(\omega)+\hat{b}_{\text{noise}}(\omega), \\
\label{eq::b_nois}
& \hat{b}_{\text{noise}}(\omega)=2\frac{\sqrt{\kappa\gamma_0}\hat{F}_{\Sigma}(\omega)}{\kappa+\gamma_{0}+\chi (\omega)-2i\omega},
\end{eqnarray}
\noindent
where $S(\omega)$ is a spectral transfer function of the memory:
\fla{
S(\omega)=\frac{\kappa-\gamma_{0}-\chi (\omega)+2i(\omega-\Delta_0)}{\kappa+\gamma_{0}+\chi (\omega)-2i(\omega-\Delta_0)},
}
$b_{\text{noise}}(\omega)$ describes the noise component in the output signal caused by the interaction with the environment modes.
\noindent

Spectral density of the output signal field $n_{\text{out}}(\omega)=\langle\hat{a}_{\text{out}}^{\dagger}(\omega)\hat{a}_{\text{out}}(\omega)\rangle$ is given by 

\begin{eqnarray}\label{eq::n_out-omega}
n_{\text{out}}(\omega)=\abs{S(\omega)}^2 n_{in}(\omega)+n_{\text{noise}}(\omega), 
\end{eqnarray}

\noindent
with the spectrum of noise photons: 

\begin{eqnarray}\label{eq::n_nois}
n_{\text{noise}}(\omega) &= \langle b_{\text{noise}}^{\dagger}(\omega)b_{\text{noise}}(\omega)\rangle= \nonumber \\
&=\mathcal{M}(\omega) n_{\text{bath}}(\omega), \label{Eq::out-spectrum}
\end{eqnarray}

\noindent
where $n_{in}(\omega)=\langle\hat{a}_{in}^{\dagger}(\omega)\hat{a}_{in}(\omega)\rangle$ is a photon number density of input signal, 
$n_{\text{bath}}(\omega)\cong n_{\text{bath}}(\omega_0)$,
the function
$\mathcal{M}(\omega)=
\frac{4\kappa [\gamma_{0}+\chi_{re}(\omega)]}{|\kappa+\gamma_{0}-2i\omega+\chi(\omega)|^2}$
reflects the effect of spectral filtering of a multi-resonator system on the appearance of quantum noise in the output radiation,
$n_{\text{bath}}(\omega_0)=1/(e^{\frac{\hbar\omega_0}{k_BT}}-1)$.
In derivation of Eq. (\ref{Eq::out-spectrum}), we take into account the same character of Langevin forces in different resonators and their rather weak dependence on frequency
$\langle \hat{F}_{n}^{\dagger}(\omega),\hat{F}_{n}(\omega)\rangle=\langle \hat{F}_{c}^{\dagger}(\omega),\hat{F}_{c}(\omega)\rangle \approx n_{\text{bath}}(\omega_0)$.

In our experiments, the input signal pulses were prepared in the coherent state of the Gaussian temporal mode: 
$\ket{\alpha}_{in}=\exp\{-\frac{1}{2}\abs{\alpha}^2+\alpha \hat{B}_g\}\ket{0}$ with an average number of photons $\abs{\alpha}^2$, where spectral properties of the signal is described by the mode operator 
$\hat{B}_g=\frac{1}{\sqrt{\sqrt{2\pi}\delta\omega_s}}\int_{-\infty}^{\infty} d\omega\exp\{-\frac{\omega^2}{4\delta\omega_s^2} \}\hat{a}_{in}^{\dagger}(\omega)$, 
$[\hat{B}_g, \hat{B}_g^{\dagger}]=1$, $\ket{0}$ is a vacuum state of waveguide modes, $\delta\omega_s$ spectral width of the input signal.
For this state of the input field, spectral density of photon number is

\begin{eqnarray}\label{eq::n_in}
 n_{\text{in}}(\omega)=\frac{\abs{\alpha}^2}{\sqrt{2\pi}\delta\omega_s}\exp\{-\frac{\omega^2}{2\delta\omega_s^2} \}.
\end{eqnarray}

\noindent
We assume that spectral width $\delta\omega_s$ covers periodic structure of resonant lines of multi-resonator quantum memory.

\begin{figure*}
     \centering
         \centering
         \includegraphics[width=0.45\linewidth]{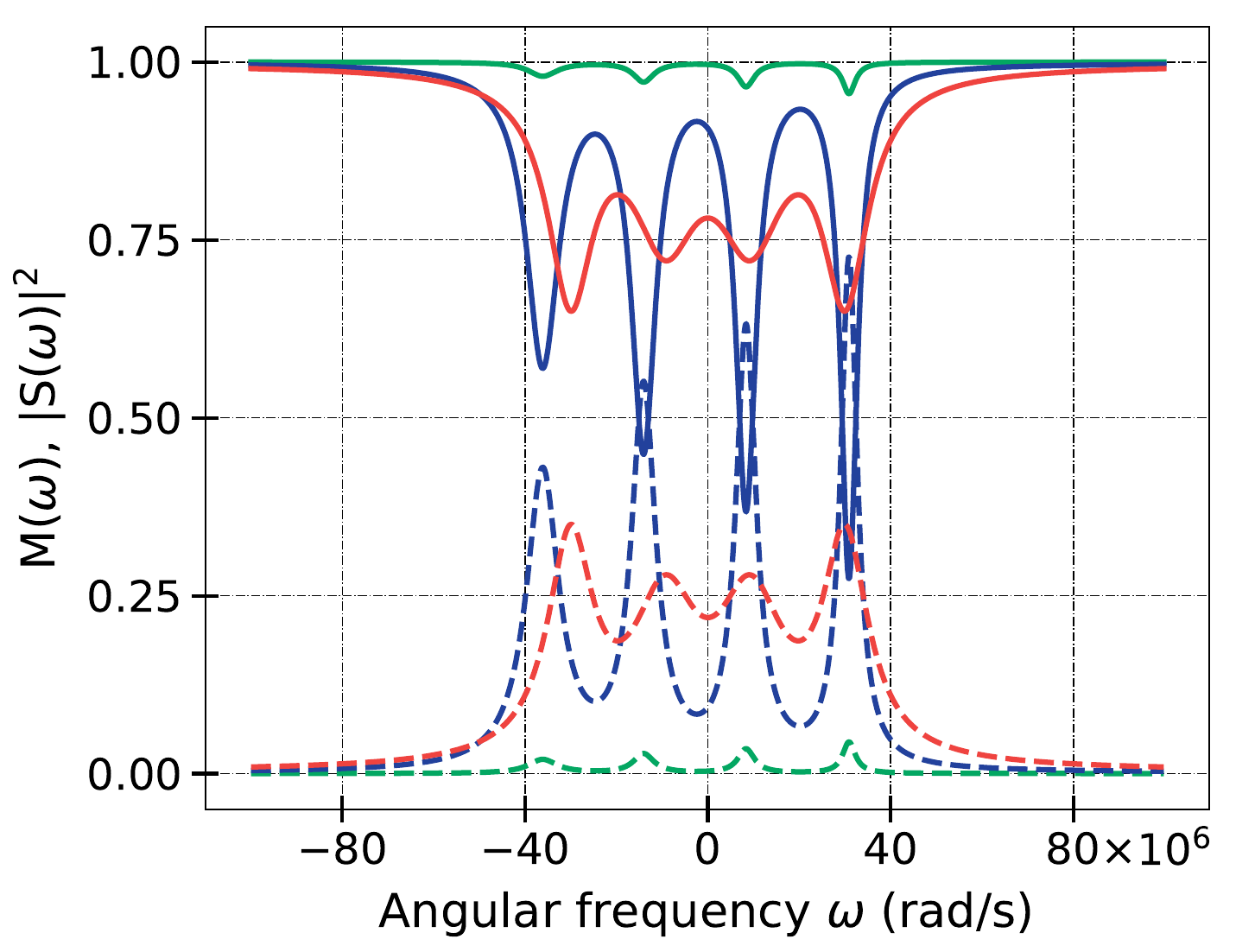}
         \caption{Filtering factor $M(\omega)$ (dashed lines) and transfer functions $|$S($\omega$)$|^2$ (solid lines). 
\textcolor{green1}{Green} curves correspond to $\gamma_{n}$=$\gamma_{0}$=6 kHz and $\Delta_{0}$=145 MHz.
\textcolor{blue1}{Blue} lines are simulations for $n_{in}<1$ where $\gamma_{n}$=$\gamma_{0}$ increases to 165 kHz.
The case of $\Delta_{0}$=0 MHz and $\kappa$ = 217 MHz is shown as \textcolor{red1}{red} curves that correspond to the fulfillment of the impedance-matching condition.}
\label{fig:spectrum}
\end{figure*}

\begin{figure*}
         \includegraphics[width=0.4\linewidth]{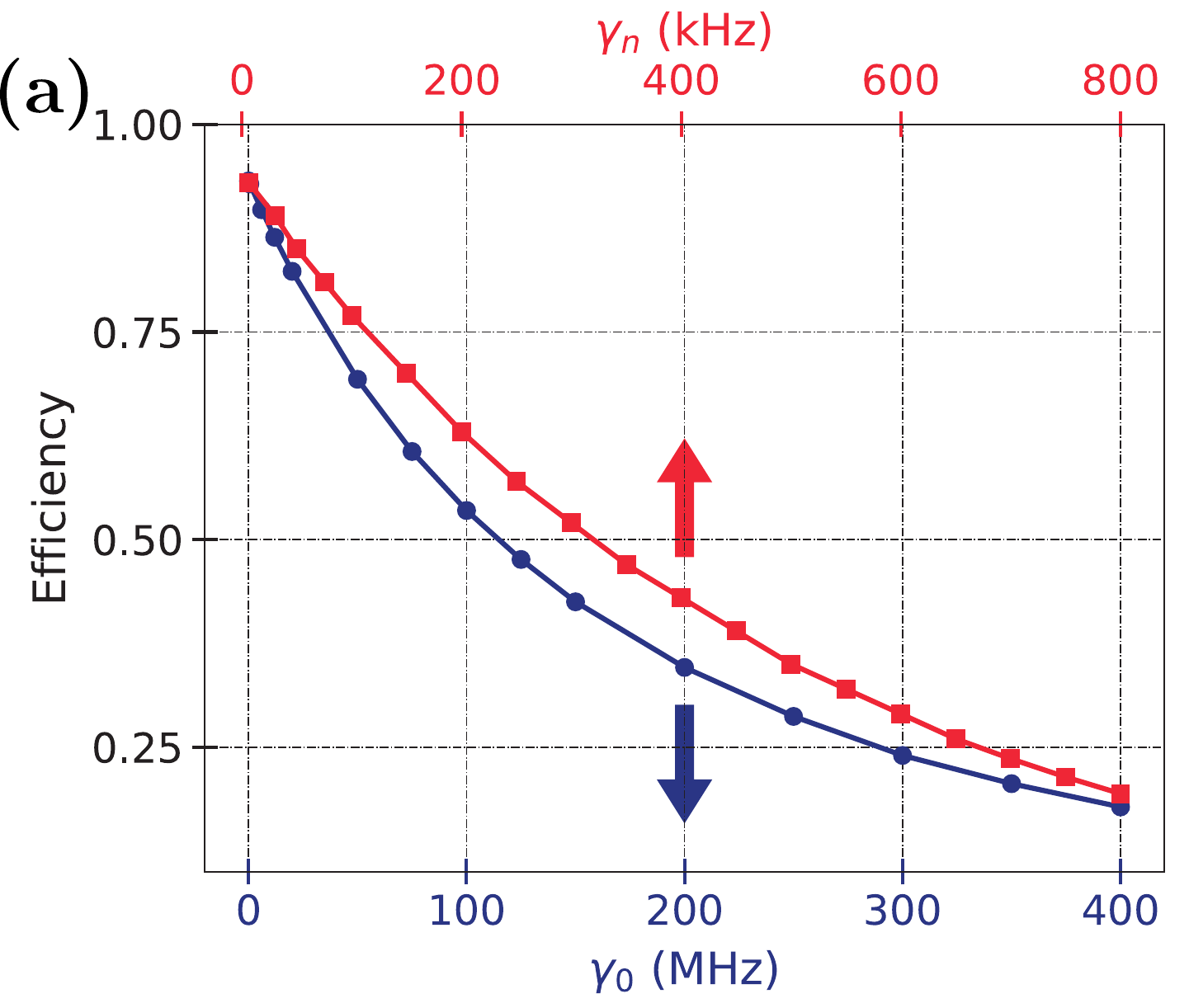}
         \includegraphics[width=0.4\linewidth]{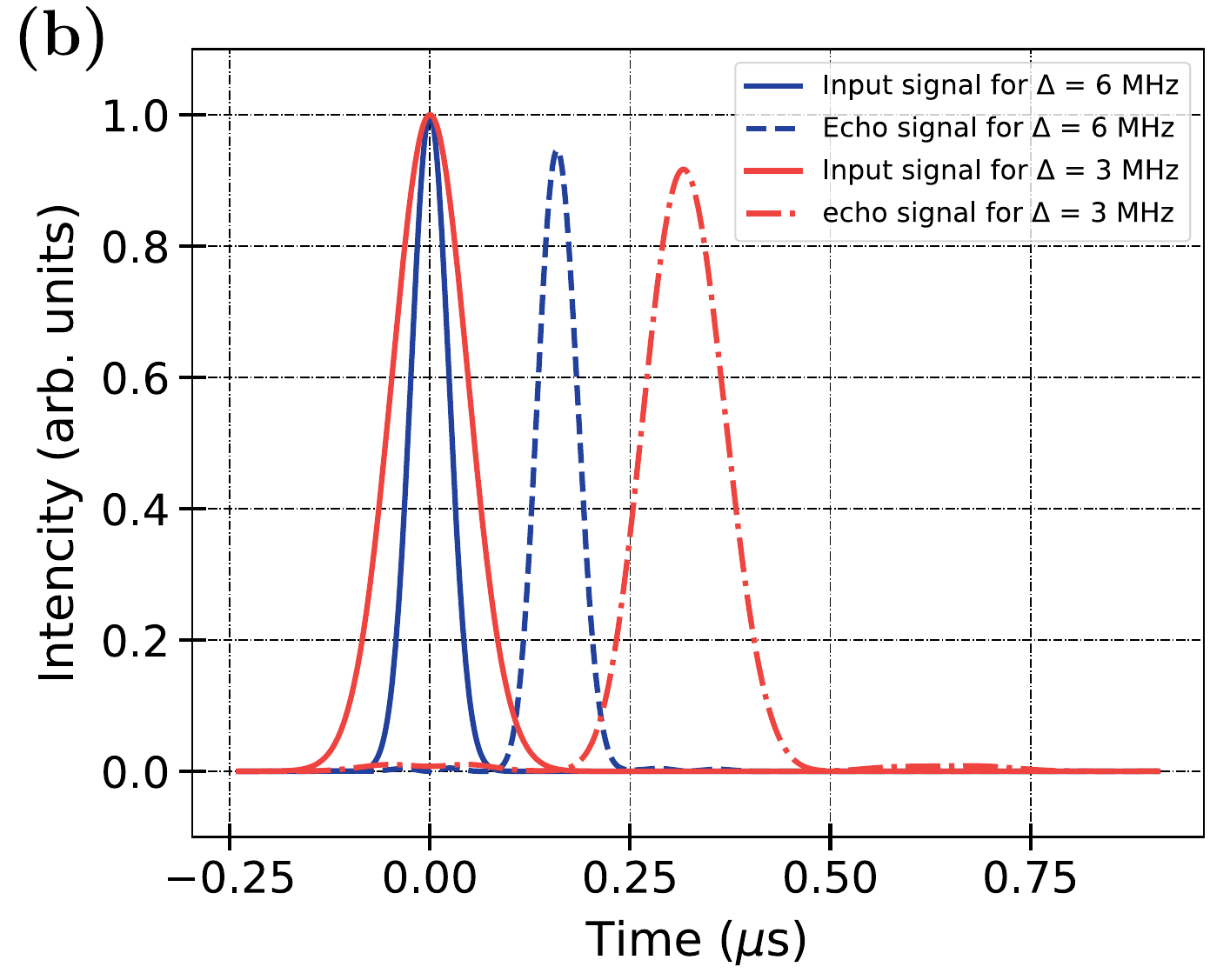}
         \caption{\textbf{(a)} Theoretical efficiency dependencies of the quantum memory on decay rates of common ($\gamma_{0}$, \textcolor{blue1}{blue} curve with circles) and internal ($\gamma_{n}$, \textcolor{red1}{red} curve with squares) resonators, respectively. Storage time is of $\sim$300 ns.
         \textbf{(b)} Demonstration of the storage of two spectral modes and the time independence of the echo signals of these modes. 
         Theoretical simulation of the echo signals of the first group of resonators with $\Delta$ = 3 MHz (\textcolor{red1}{red} curves) and  the second group of resonators with $\Delta$ = 6 MHz (\textcolor{blue1}{blue} curves).
         The optimized parameters to meet the impedance matching condition are
         $\kappa$ = 280 MHz, $\gamma_{n} = \gamma_{0}$ = 6 kHz, $g_{\Delta=3}$ = 11.57 MHz, $g_{\Delta=6}$ = 16.36 MHz.
         }
         \label{fig:eff}
         % \label{fig:Delta6and3}
\end{figure*}

\begin{figure*}
\centering
\includegraphics[width=0.4\linewidth]{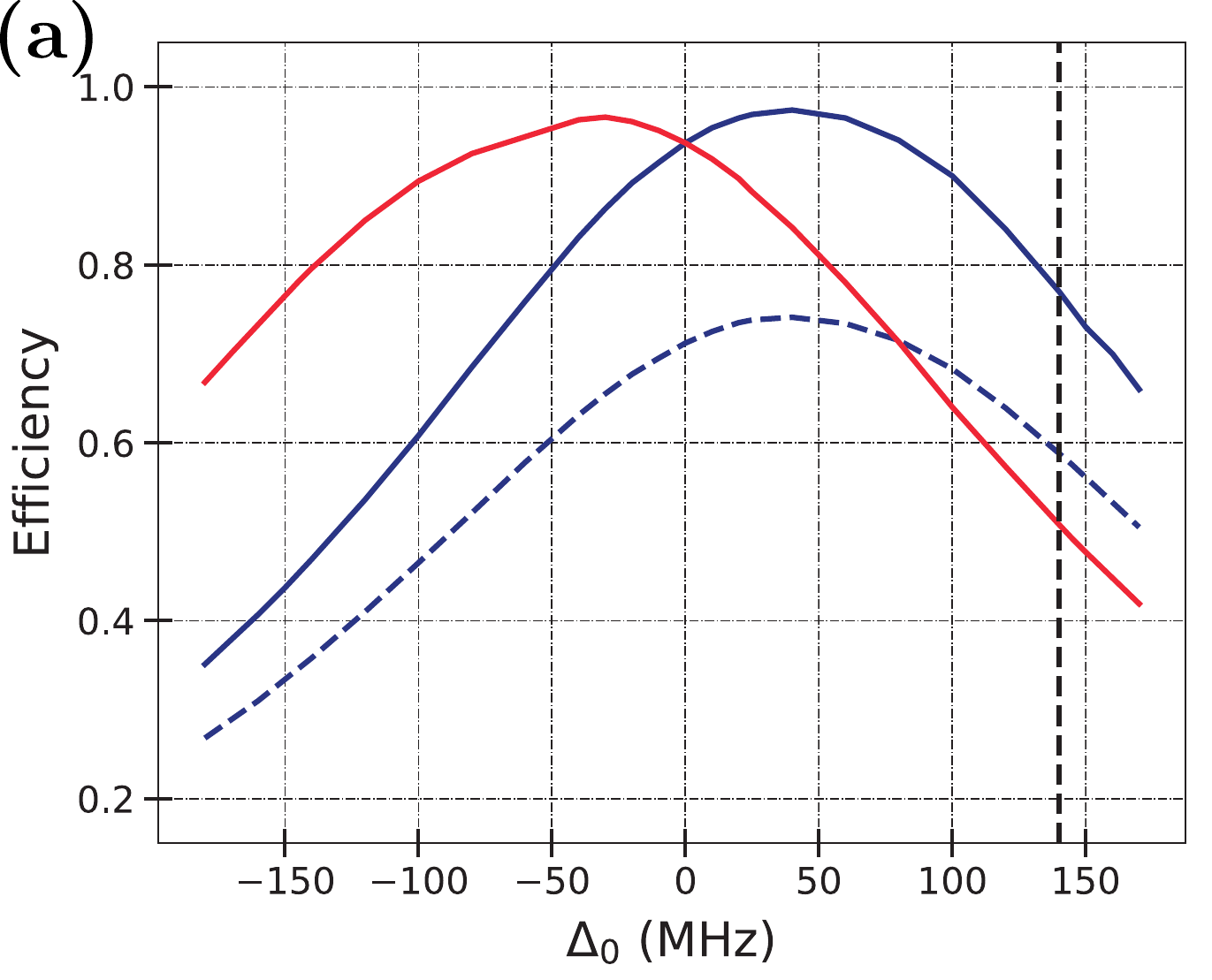} 
\includegraphics[width=0.4\linewidth]{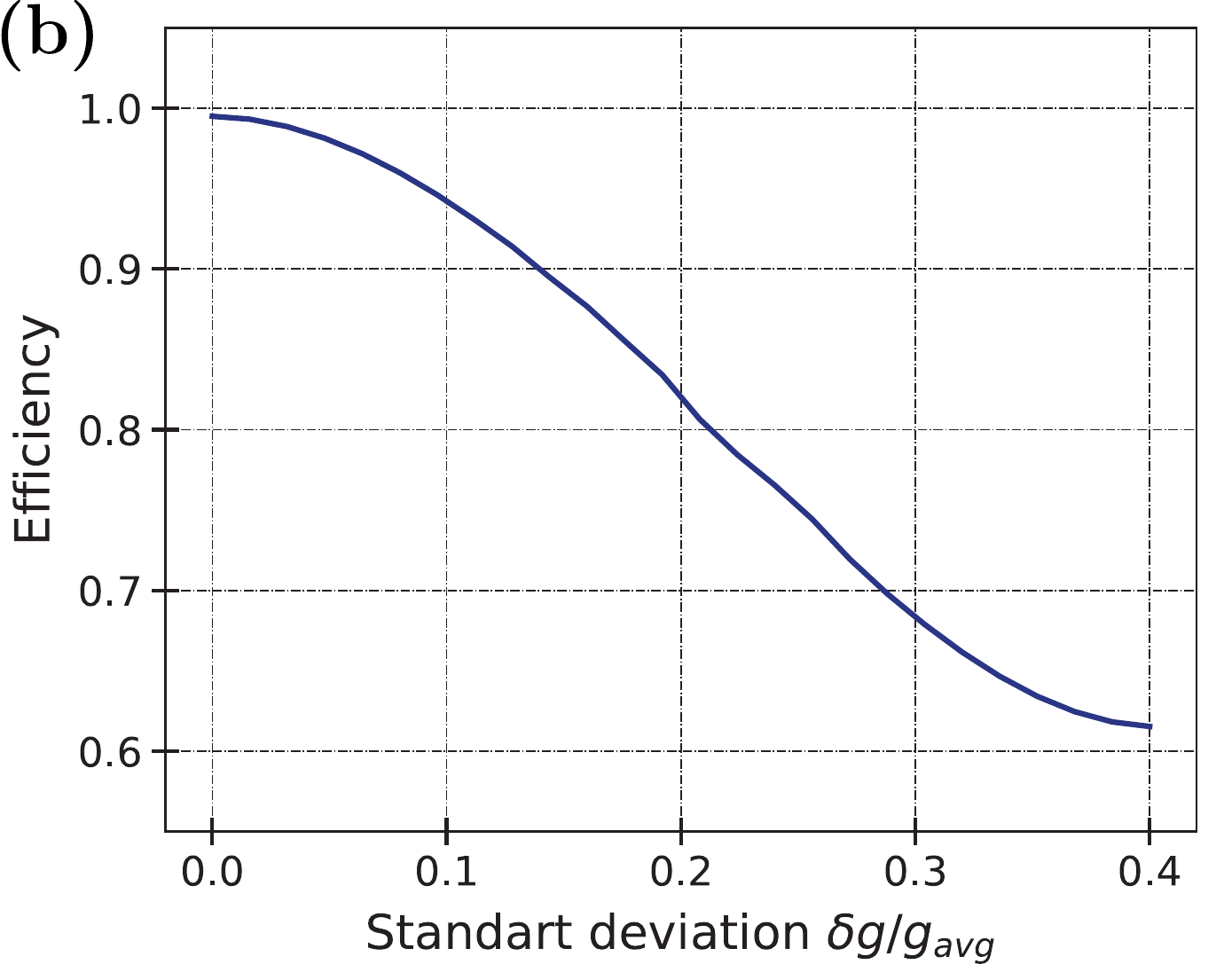}
\caption{
\textbf{(a)} Dependence of the efficiency on the common resonator frequency shift. 
The \textcolor{blue1}{blue} and \textcolor{red1}{red} solid lines correspond to the first and second groups of resonators, respectively, in the case of bright pulses. 
The \textcolor{blue1}{blue} dashed line is the dependence for the first group of resonators, the case of single photon pulses.
The black dashed vertical line show the parameters corresponding to the experiment.
The dependence maxima correspond to the case of complete impedance matching.
\textbf{(b)} Dependence of the efficiency on relative statistical uncertainty of the coupling constant between the internal and the common resonators.
}
\label{fig:non-ideal}
\end{figure*}

By using Eqs. (\ref{eq::in-out}),(\ref{eq::n_nois}) for $\abs{S(\omega)}^2$ and $\mathcal{M}(\omega)$, we analyze the spectral properties of efficiency and the level of noise suppression in the working spectral range covering the four frequencies of each resonator group.
As it was shown in \cite{Perminov2019} an almost unit efficiency is possible 
for the number of resonators $N=4$ at the impedance matching condition 
\fla{
\kappa = \gamma_0 + 2\pi g_{n}^{2}/(\gamma_n/2+\Delta),
}
where 
$\Delta_0=0$, $\Delta_n =  (n+1/2)\Delta$, ($n\in\{...,-2,-1,0,1,...\}$) and
$\gamma_0$, $\gamma_n$ are much smaller than $2\pi g_{n}^{2}/\Delta$ with $\Delta$ being frequency spacing between nearest internal resonators.
Using the experimental parameters of the fabricated quantum memory, we calculate the behavior of the functions $\abs{S(\omega)}^2$ and $\mathcal{M}(\omega)$, as it is depicted in Fig. \ref{fig:spectrum}.

We show that the quantum memory chip has filtering properties and can significantly suppress quantum noise present in the environment. 
The suppression factors  are $\sim$0.14 and 0.006 for $\gamma_{0}$ = $\gamma_{n}$ = 165 and 6 kHz, respectively. 
This means that the demonstrated quantum memory chip can operate at higher temperatures than 10 mK. 
For the single-photon mode and the desired signal-to-noise ratio of $\sim$100, the temperatures are 100 and 600 mK for $\gamma_{0}$ = $\gamma_{n}$ = 165 and 6 kHz, respectively.

The experimentally measured behavior of the output signal from the quantum memory is given by 
\fla{
n(t) = \langle \hat{a}_{out}^{\dagger}(t) \hat{a}_{out}(t) \rangle,
\label{eq::n_out}
}
where the Fourier-transformed operator $\hat{a}_{out}(t)$ is  
\begin{eqnarray}
\hat{a}_{out}(t) = \frac{1}{\sqrt{2\pi}}\int d\omega e^{-i\omega t }S(\omega) \hat{a}_{in}(\omega).
\label{Time_domain-solution}
\end{eqnarray}

\noindent
Eq. \eqref{eq::n_out} is used for theoretical fit of experimental data for intensive and single photon level microwave pulse. 
We plot theoretical temporal behavior of the output signal using
Eq. \eqref{Time_domain-solution} with experimental parameters $\Delta_0$, $\kappa$, $\gamma_0,\gamma_n$ and $g_n$ in Fig. 2a and 2b of the main text. 

For the same experimental parameters we separately plot the dependence of the efficiency on the decay constants $\gamma_0$ and $\gamma_n$ in Fig. \ref{fig:eff}\textbf{a}.
The decay constant of the common resonator $\gamma_0$ affects the efficiency approximately 500 times weaker than the decay in the internal resonator $\gamma_n$.
We explain the weak influence of internal losses in the common resonator by the presence of a large constant coupling of this resonator with an external waveguide such that the field does not spend much time there.
Thus, we conclude that the effect of saturable two-level systems are observable mostly in the internal resonator. 

We show that different frequency spacing of two group of resonators with impedance matching being fulfilled in both of them. 
In Fig. \ref{fig:eff}\textbf{b} we present theoretical simulation of the echo signals of the first group of resonators with $\Delta_{1-4}=3$  MHz and the second group of resonators with $\Delta_{5-8}$ = 6 MHz. 
The central frequencies of the first and second groups are shifted from the central frequency of the common resonator by -7 and 16 MHz, respectively. 
At optimal parameters of the quantum memory, both modes are experiencing the same efficiency with their recall time being separate by about a pulse width.

The fabricated multiresonator may have a deviation of its parameters from the designed values. 
We assess how a deviation in the resonant frequency of the common resonant may effect the efficiency of the memory.
In Fig. \ref{fig:non-ideal}\textbf{a} we show the dependence of the power efficiency on the value of the $\Delta_0$ for the first and the second memory units at the experimental conditions. 
At $\Delta_0=0$ the same efficiency occurs for both groups of resonators. 
At small deviation of $\Delta_0$ ( $\abs{\Delta_0} < N g$ with $N$ being the number of resonators in each group) there is a small increase of efficiency in one of the groups with corresponding decrease in another depending towards which group the common resonator's frequency is shifted. 
However at large detuning $\abs{\Delta_0} > 4 g$ both groups lack the impedance matching with a significant decrease in efficiency.

Also we perform Monte-Carlo-like simulation for nonuniform coupling between the internal resonator and the common resonator. 
For each simulation we sample $g_j$ from normal distribution with an average $g_{\text{avg}}$ and variance $\delta g^2$ and  numerically solve Eqs. \eqref{eq1}-\eqref{eq::in-out-time}.
We average the power efficiency of the first echo over 100 simulations.   
The resulted dependence of the efficiency on the coupling variance is presented in Fig. \ref{fig:non-ideal}\textbf{b}.
For highly efficient storage the root mean square uncertainty of the coupling constants has to be smaller than $\delta g / g_{\text{avg}} < 0.05$

\begin{figure*}[htp]
\hspace{0.18\linewidth}
\includegraphics[width=0.17\linewidth]{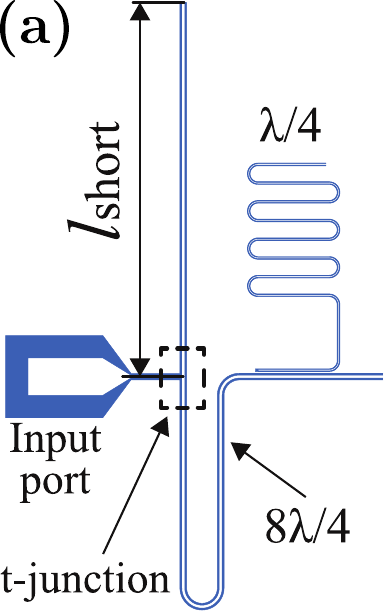}
\hspace{0.05\linewidth}
\includegraphics[width=0.4\linewidth]{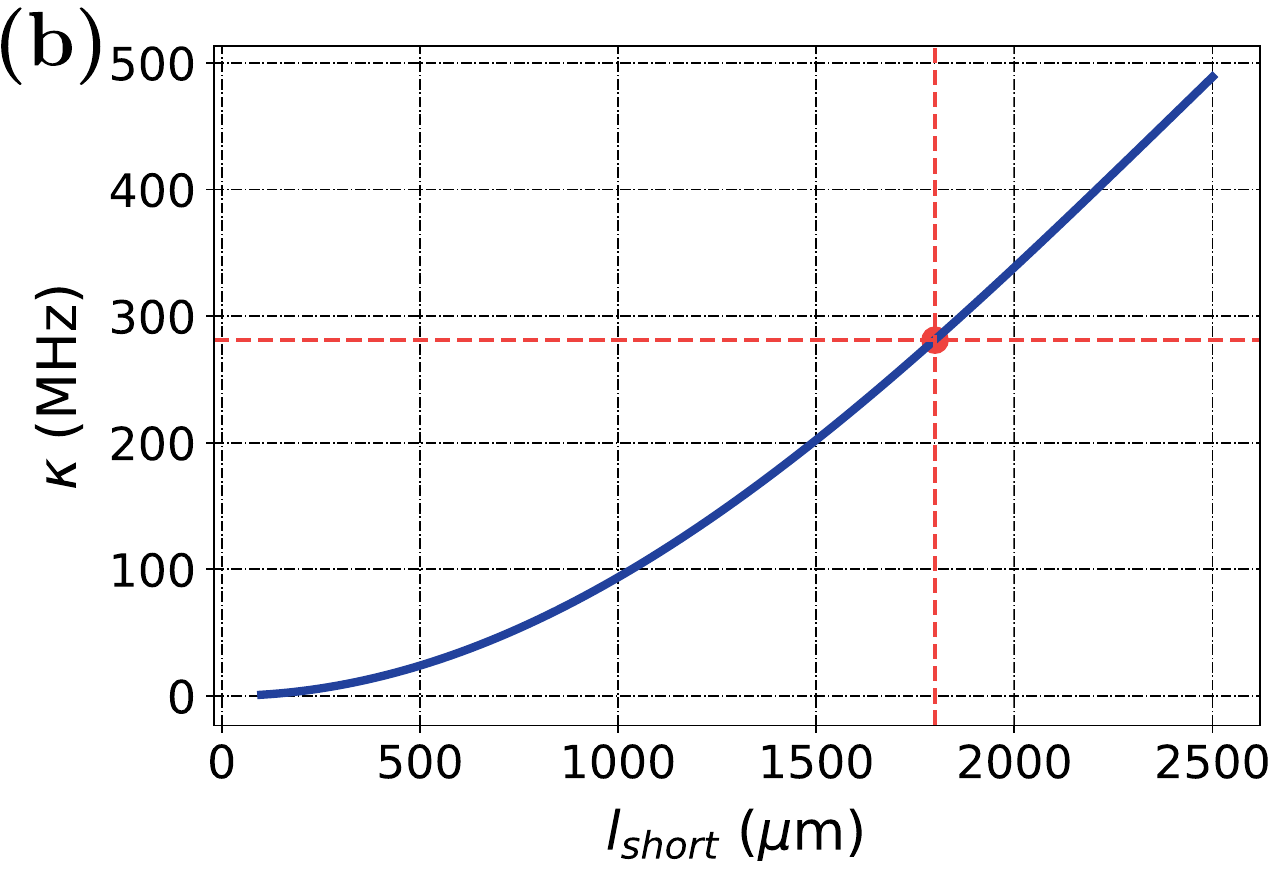}
\includegraphics[width=0.4\linewidth]{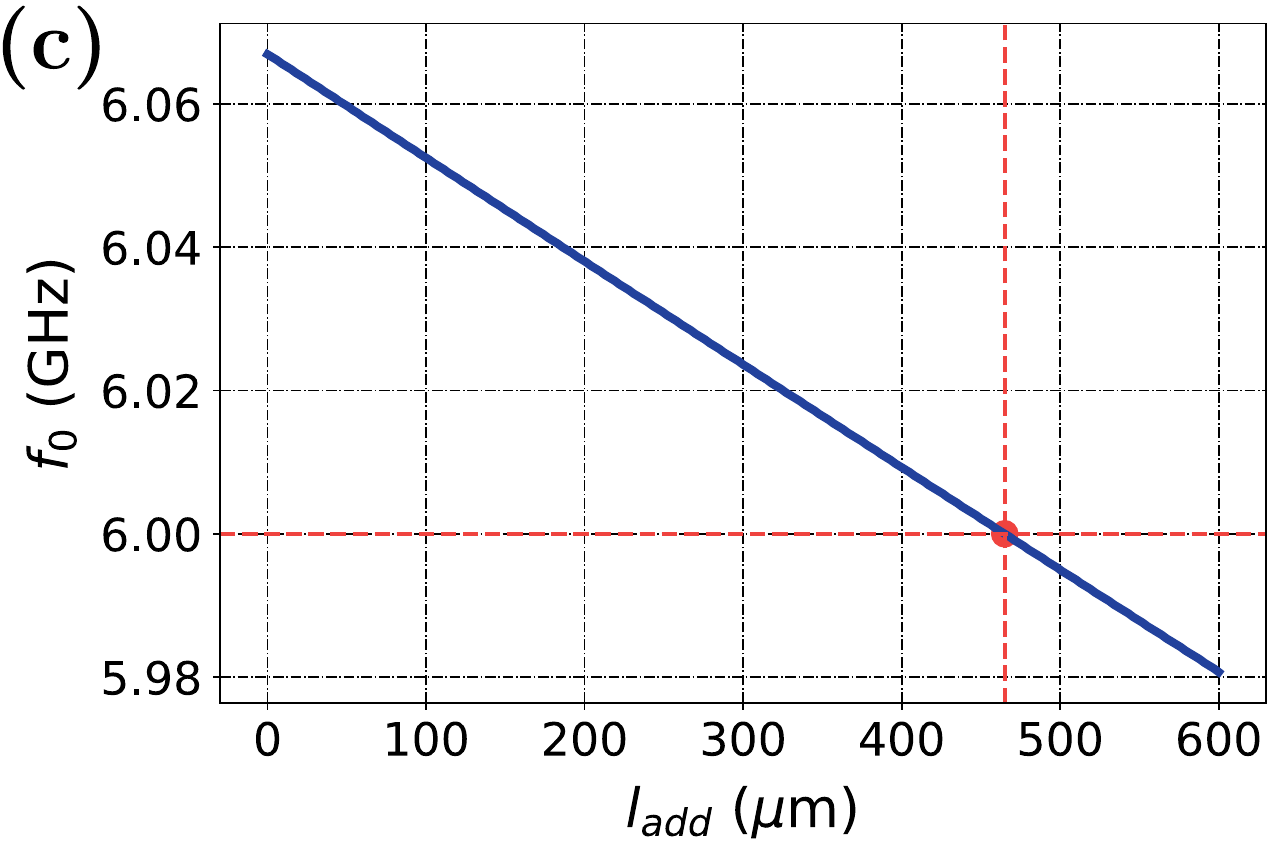}
\includegraphics[width=0.4\linewidth]{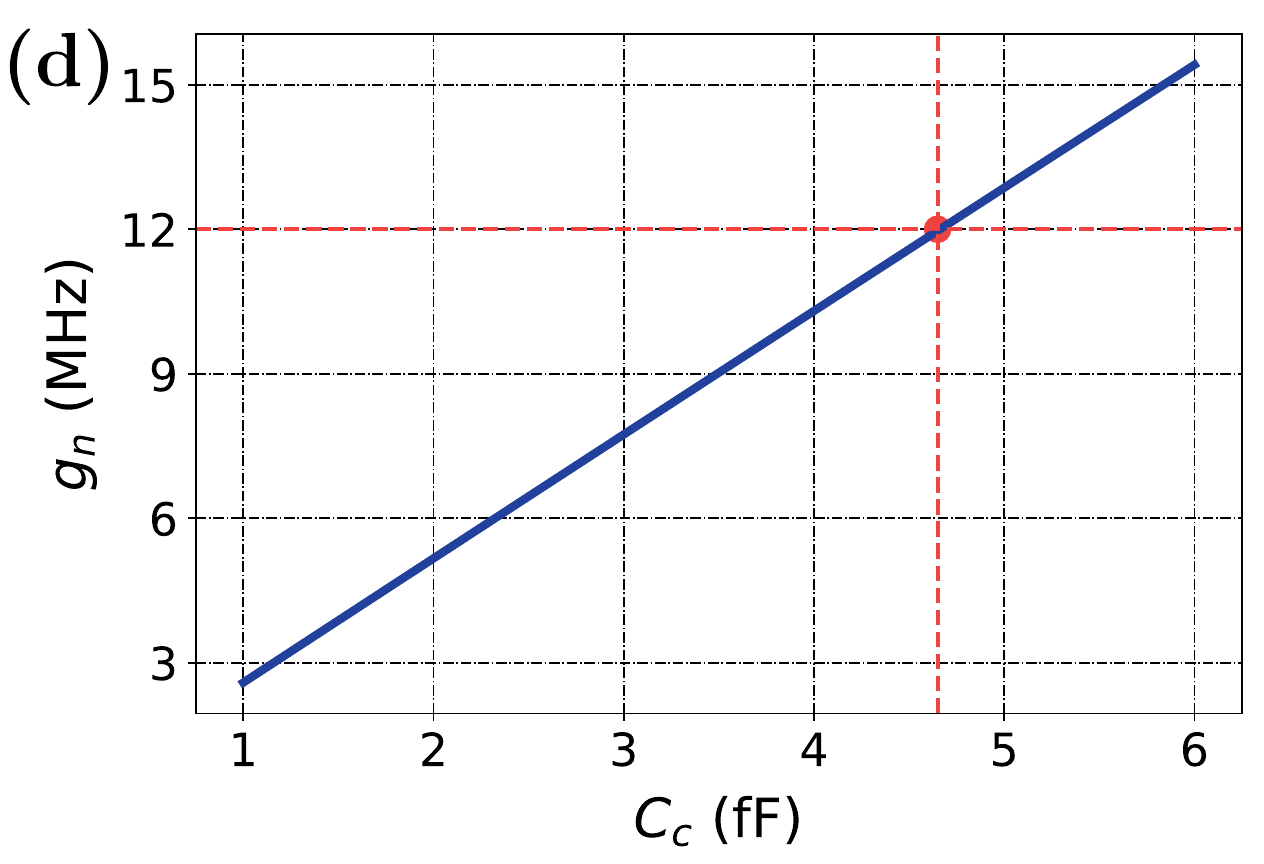}
\caption{\textbf{a)} The input part of the common $8\cdot\lambda/4$ resonator for clarification of $l_{\text{short}}$ parameter and t-junction
\textbf{b)} The dependence of the coupling constant $\kappa$ on the length $l_{\text{short}}$. \textcolor{red1}{Red} dot is the desired point with $\kappa=281$ MHz and length $l_{\text{short}}=1800\;\mu m$
\textbf{c)} The dependence of resonant frequency of the common resonator $f_0$ on its additional length $l_{\text{add}}$. \textcolor{red1}{Red} dot is the desired point with $f_0=6$ GHz at $l_{\text{add}} = 465\;\mu m$
\textbf{d)} The dependence of the coupling strength $g_n$ on the coupling capacitance $C_c$ between the common and the internal resonators. \textcolor{red1}{Red} dot is the desired point with $g_n=12$ MHz at coupling capacitance of $C_c=4.65$fF.
}
\label{fig:simulations}
\end{figure*}

\section{Design and fabrication of the quantum memory chip}\label{app:B}

\subsection{Calculation of circuit parameters}

\begin{figure}[htp]
\includegraphics[width=0.9\linewidth]{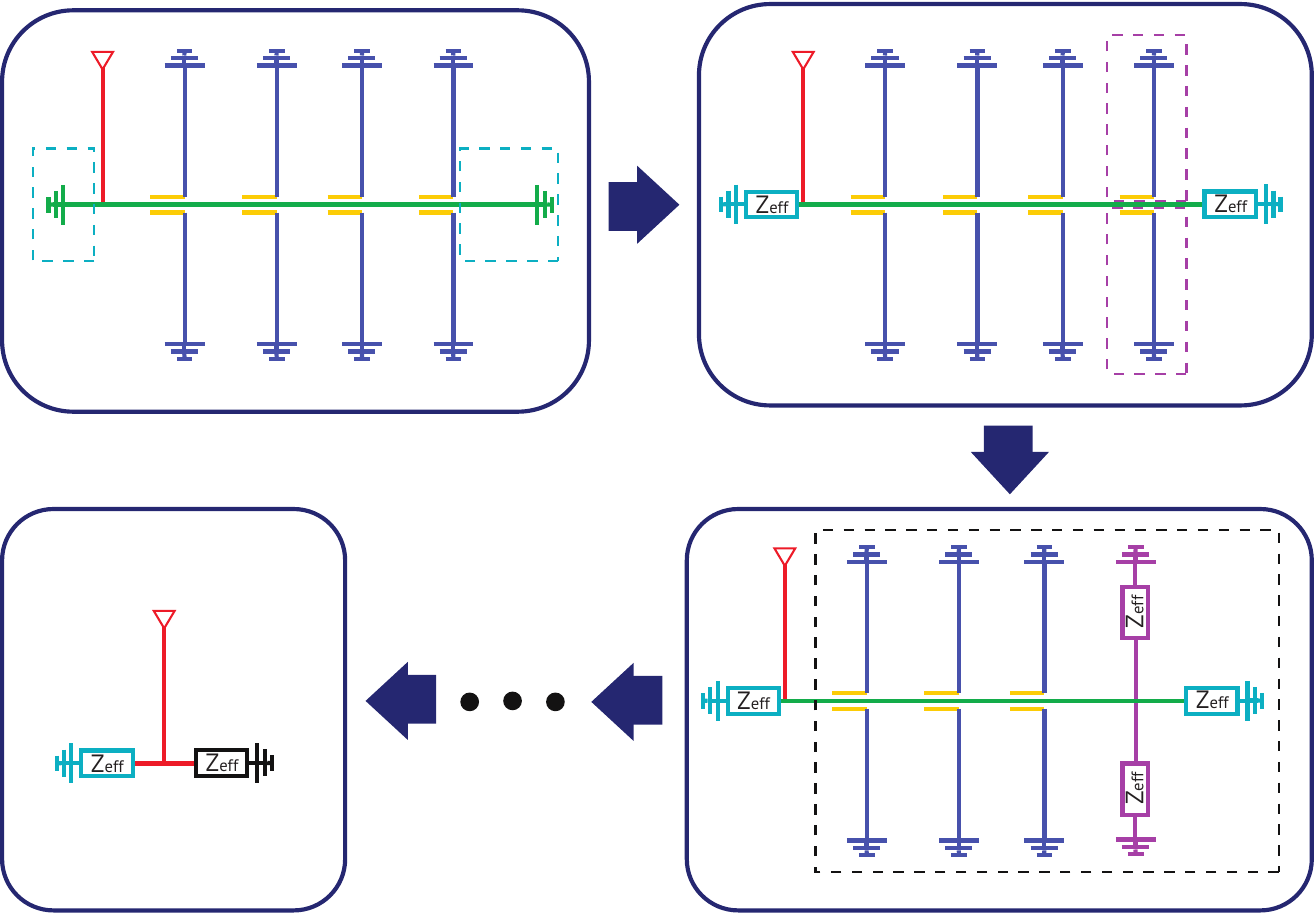}
\caption{The concept of the iteratively reduction method.}
\label{fig:CalculatingMethod.eps}
\end{figure}

For high efficient quantum memory it is necessary to achieve the impedance matching condition that relates the coupling strength $g_{n}$ between eight internal $\lambda/4$ and common $8\cdot\lambda/4$ resonators, coupling $\kappa$ between the input waveguide of the common $8\cdot\lambda/4$ resonator and frequency spacing $\Delta$. 
Therefore, it is crucial to have a method for calculating these key parameters with high accuracy. We use the method of sequential reduction of long transmission lines with distributed parameters to concentrated elements for calculating quantum memory parameters \cite{pozar2011}. 
The method iteratively reduces the circuit's elements to an effective impedance as it is illustrated in Fig. \ref{fig:CalculatingMethod.eps}. 
On each iteration, the effective impedance is calculated as (\ref{eq::Z_eff})-(\ref{eq::Refl_Coeff})

\begin{gather}
    \label{eq::Z_eff}
    Z_{\text{eff}}=Z_0\cdot\frac{(1+R\cdot e^{-2\cdot(\alpha+j\beta)\cdot l})}{(1-R\cdot e^{-2\cdot(\alpha+j\beta)\cdot l})}, \\
    \label{eq::Refl_Coeff}
    R=\frac{Z_{\text{eff}}-Z_0}{Z_{\text{eff}}+Z_0},
\end{gather}

\noindent
where $\alpha+j\beta$ is the complex propagation constant that depends on relative permittivity of the substrate (for Si substrate $\epsilon=11.9$) and the gap-conductor-gap ratio (10-16-10 $\mu$m and 4-7-4 $\mu$m for common and internal resonators, respectively), $l$ is the length and $Z_0$  ($\approx$ 50 Ohm for our choice of gap-conductor-gap ratios) is the characteristic impedance of the distributed element.

\subsection{Common resonator coupling constant $\kappa$ calculation}

\begin{table*}
%\begin{ruledtabular}
\begin{tabular}{ |c|c|c|c| }
\hline
\text{Resonator number}&
\text{Designed frequency, GHz}&
\text{Measured frequency, GHz}&
\text{Error, \%}\\
\colrule
1 & 5.9895 & 5.9383 & 0.85\\
2 & 5.9925 & 5.9418 & 0.84\\
3 & 5.9955 & 5.9453 & 0.84\\
4 & 5.9985 & 5.9488 & 0.83\\
5 & 6.0015 & 5.9502 & 0.85\\
6 & 6.0045 & 5.9532 & 0.85\\
7 & 6.0075 & 5.9562 & 0.85\\
8 & 6.0105 & 5.9592 & 0.85\\
\hline
\end{tabular}
%\end{ruledtabular}
\caption{\label{tab:table1}%
Parameters of the internal resonators.
}
\end{table*}
%\twocolumngrid

For the common resonator we choose $8\cdot\lambda/4$ design  since it does not need a large capacitance at the input of the circuit as in $9\cdot\lambda/4$ resonator, where it is difficult to precisely determine this capacitance. 
Moreover, $8\cdot\lambda/4$ resonator has the advantage of input-output port being coupled to the common resonator using voltage tap. 
Hence the coupling constant $\kappa$ is set simply by the length $l_{\text{short}}$ of the section between grounded end and input-output port (Fig. \ref{fig:simulations}\textbf{a}) that can be calculated with relatively high precision. 
We model the common resonator as a parallel LC-circuit with effective capacitance $C$ and inductance $L$, that are numerically calculated by the  sequential reduction method. 

We express the coupling constant $\kappa$ as

\begin{equation}
    \label{eq::kappa}
    \kappa=\frac{\omega_{0}}{Q_c}=\frac{\omega_{0}\cdot Z_{LC}}{Z_0},
\end{equation}

\noindent
where $\omega_{0}$ is the angular frequency of the common resonator, $Z_0$ is the characteristic impedance, and $Z_{LC}=\sqrt{\frac{L}{C}}$ is the characteristic impedance of equivalent parallel LC-circuit with effective capacitance $C$ and inductance $L$. 
The numerically calculated dependence of the coupling constant $\kappa$ on $l_{\text{short}}$ is shown in Fig. \ref{fig:simulations}\textbf{b}. 
We choose $l_{\text{short}}$ to be $1800\; \mu m$ that
provides the desired value of coupling constant $\kappa\approx 281$ MHz.

\subsection{Extra length $l_{\text{add}}$ of the common $8\cdot\lambda/4$ resonator}

The presence of the t-junction at the input of the common resonator (see Fig. \ref{fig:simulations}\textbf{a}) shifts its resonant frequency. This shift has to be adjusted by adding an extra length $l_{\text{add}}$. The resonant frequency of the common resonator is calculated as a resonance frequency of equivalent parallel LC-circuit, which parameters are calculated by the sequential reduction method. 
The dependence of the common resonator's frequency $f_0$ on an extra length $l_{\text{add}}$ is shown in Fig. \ref{fig:simulations}\textbf{c}. 
For the desired value of resonant frequency $f_0$ of $6$ GHz, we choose $l_{\text{add}}$ to be $465\;\mu m$.

\subsection{Coupling strength $g_n$ between the internal $\lambda/4$ and the common $8\cdot\lambda/4$ resonators}

The coupling strength $g_n$ between internal $\lambda/4$ and common $8\cdot\lambda/4$ resonators is determined by coupling capacitance and the their relative position. 
%For its evaluation it is necessary to replace internal $\lambda/4$ and common $8\cdot\lambda/4$ resonators by parallel LC circuits with effective capacitancies $C_{0}$, $C_{n}$ and inductance $L_{0}$, $L_{n}$ using the step-by-step reduction method, described above. 
For two resonators, that are represented as an effective LC-circuits by the sequential reduction method, the $g_n$ in units of angular frequency is \cite{chen2018Thesis}:

\begin{equation}
    \label{eq::g_n}
    g_n=\frac{1}{2}\cdot\frac{C_c}{\sqrt{(C_{n}+C_c)\cdot(C_{0}+C_c)}}\cdot\sqrt{\omega_{0}\cdot\omega_{n}},
\end{equation}

\noindent
where $C_c$ is the coupling capacitance between common and internal resonators, $\omega_{0}$ and $\omega_{n}$ are the angular frequencies of the common resonator and the internal resonator respectively, $C_{0}$ and $C_{n}$ are the capacitance of equivalent parallel LC-circuit for common and internal resonators, respectively. The dependence of the coupling strength $g_n$ on the coupling capacitance $C_c$ between common and internal resonators is shown in Fig.\ref{fig:simulations}\textbf{d}. 
For coupling constant $g_n$ of $12$ MHz, we choose the coupling capacitance $C_c$ between common and internal resonators to be $4.65$ fF.

%\begin{figure}[htp]
%\includegraphics[width=\linewidth]{Pictures/gCalc.eps}
%\caption{The dependence of the coupling strength $g_n$ on the coupling capacitance $C_c$ between common and internal resonators. Red dot is the desired point with $g_n$ equals to $2\pi\cdot12$ MHz and coupling capacitance $C_c$ between common and internal resonators equals to $4.65$ fF}
%\label{fig:gCalc.eps}
%\end{figure}

\subsection{Device fabrication}
The device is fabricated on high-resistivity silicon substrate ($\rho>10000$   Ohm · cm, 525 $\mu$m
thick). Firstly, the substrate is cleaned with
a Piranha solution at 80°C, followed by dipping in 2$\%$ hydrofluoric bath. Then 100 nm aluminum film is deposited using e-beam evaporation
in a ultra-high vacuum deposition system. After that, 600 nm thick positive photoresist is spin
coated. Resonators and ground plane are defined using a laser direct-writing lithography system, and then wet-etched using commercial Al etchant solution. The photoresist is stripped in N-methyl-2-pyrrolidone at 80°C for 3 hours and rinsed in IPA (isopropyl alcohol) with sonication. 

Finally, aluminum free-standing crossovers are fabricated for the suppression of parasitic modes. The 3 $\mu$m thick sacrificial photoresist layer is patterned and reflowed at 140°C. An aluminium layer of 300 nm is then evaporated with an in-situ Ar ion milling to remove the native oxide. 
The second layer of 3 $\mu$m photoresist is used as a protective mask, and the excess metal is dry etched in inductively coupled plasma. 
A damaged layer of
the photoresist is then removed in oxygen plasma, and both photoresist layers are stripped in N-methyl-2-pyrrolidone at 80°C.

The difference between designed and measured internal resonator frequencies listed in Table 1.

\section{Details on Experimental setup}
\label{app:C}
\begin{figure*}[htp]
\includegraphics[width=0.8\linewidth]{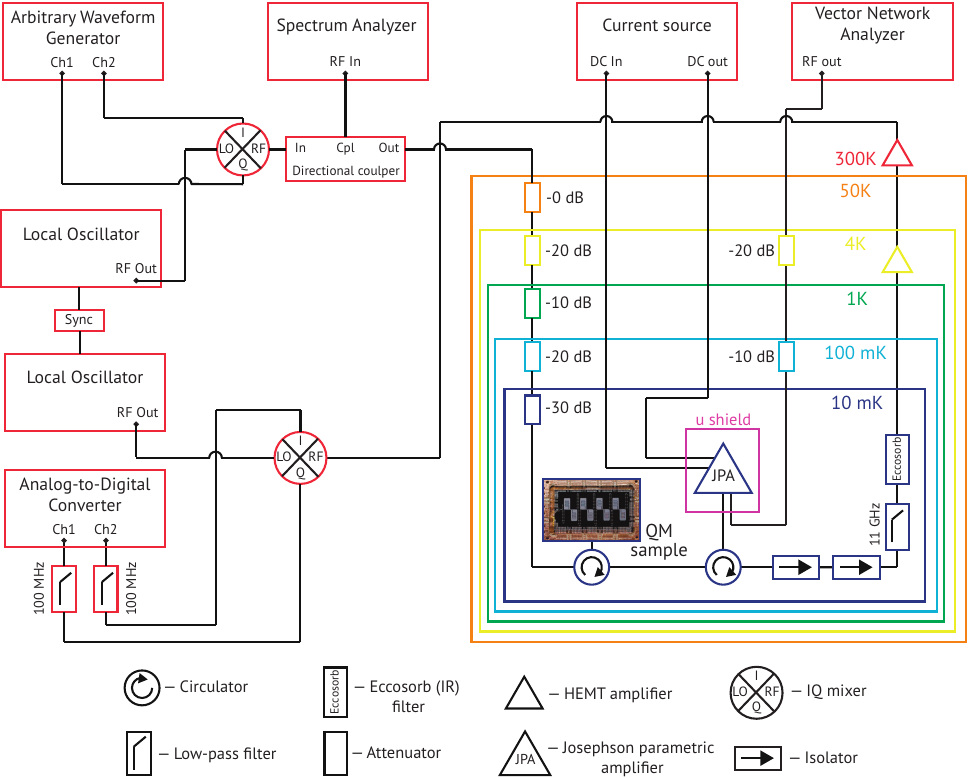}
\caption{Description of the electrical and cryogenic setup of the experiment. The colors indicate the different temperature stages for the components.}
\label{fig:ExperimentalSetup.eps}
\end{figure*}

The detailed experimental setup is shown in Fig. \ref{fig:ExperimentalSetup.eps}. Most of the control electronics are located at room-temperature with some components of the experimental setup being  
distributed over different cooling stage of the $^3$He--$^4$He dilution refrigerator as illustratively indicated in Fig. \ref{fig:ExperimentalSetup.eps} 

The probe pulse is generated using two channels of an arbitrary waveform generator  at 2.4 GSPS that drives I and Q channels of IQ mixer for single-sideband upconversion to the radio frequency.
The Spectrum Analyzer is used for calibration of single-sideband upconversion and rejecting image sideband. 
Before reaching the quantum memory, the probe is consequently attenuated by 20 dB at 4 K, 10 dB at 1 K, 20 dB at 100 mK and 30 dB at 10 mK in an $^3$He–$^4$He dilution cryostat.

After passing through the quantum memory the probe is amplified by a narrowband quantum-limited Josephson parametric amplifier (JPA) with a gain of $G_{\text{JPA}}=20$ dB, the bandwidth of 40 MHz and input 1 dB-compression power of -115 dBm.
JPA's central frequency is tunable from 5.5 GHz to 6.5 GHz by applying DC current bias from a precision current source to the coil located around the sample holder.
JPA is pumped by vector network analyzer with a twice of the central frequency.

After the JPA, the probe pulse is guided through a low-pass filter with cut-off frequency of 11 GHz and infrared-photon blocking custom-built filter (Eccosorb) at 10 mK and is amplified with a low noise amplifiers based on a high-electron-mobility transistor (HEMT) at 4 K and room temperature stages. The RF amplification is followed by down-conversion to an intermediate frequency that is further low-pass filtered for suppression of out-of-band noise and standing waves. The signal is further sampled by an analog-to-digital converter at 500 MSPS.
Two different pre-synchronized local oscillators with a tunable relative phase are used for up- and down-conversion to measure different quadratures for quantum process tomography experiments.

\section{Details on quantum process tomography } \label{app:D}
Coherent state quantum process tomography analyzes a quantum process with respect to its action on a set of coherent states. 
Thus, it is required to have an ability to connect and disconnect the device under test from the circuit with an input of these coherent states. 
In our case, we effectively ``disconnect" the quantum memory from the circuit by sending a microwave signal, that we call reference, out of resonance by 105 MHz.

In this regime, we first calibrate the reference set of coherent states.  
For this purpose, we apply the method of the measured amplitude moments \cite{Eichler2011} that allows determining the value of the probe's amplitude at the input of the amplifier while operating even with a noisy HEMT-based amplifier.
Next we calibrate the efficiency of the homodyne detector by performing quadrature measurement with the JPA at the same amplitudes as in the previous experiment. 
The individual quadrature $Q$ is measured by integrating continuous output $i(t)$ from a single channel of the IQ-mixer over the waveform of the chosen mode at given phase difference between the local oscillators:
\fla{
Q = \int i(t) \psi(t) ,
}
where integration occurs over a window of 1.5 $\mu$s with $\psi(t)$ being normalized Gaussian waveform with full width half maximum of 115 ns.
By reconstructing the density matrices of the input coherent states with $2\cdot10^5$ measured quadratures values by maximum likelihood method \cite{Lvovsky_2004} we adjust the efficiency of the homodyne detector such that reconstructed amplitudes match the known amplitudes of the probes. 
%The resulted efficiency of $48\pm3$\% is used for further experiments on quantum process tomography.   

The narrowband nature of the JPA demands us to set JPA gain separately for the reference and the probe that is in resonance with the quantum memory and is used for quantum process tomography. 
A large number of microwave components with a slight deviation of their characteristic impedances from 50 Ohm  leads to undesired reflections and unequal reflection coefficient $S_{11}$ at different frequencies in a complete experimental setup. 
Hence, different gains are required for the reference and the probe to compensate for this variation. 
For a fixed value of JPA's gain for the reference, we adjust JPA's gain for the probe such that the total integrated over 1.5 $\mu$s energy of the probe after the quantum memory is $\sim$96 $\%$ compared to the energy of the reference. Both probe and reference signals are sent with large intensities. The resulted gains are equal to 19.6 dB and 20.3 dB for the reference and probe signals, respectively.

\bibliography{QRAM}

\end{document}